\documentclass[aps,preprint,pre]{revtex4}
\usepackage{epsfig}
\usepackage{graphicx}
\begin{document}
\title{Classical Scattering for a driven inverted Gaussian potential in terms
of the chaotic invariant set}
\author{A. Emmanouilidou$^{1}$, C. Jung$^{2}$, L. E. Reichl$^{3}$}
\affiliation{$^{1}$Max Planck Institute for the Physics of Complex Systems,
N$\ddot{o}$thnitzer Stra$\beta$e 38, 01187 Dresden, Germany }
\affiliation{$^{2}$Centro de Ciencias Fisicas, UNAM, Apdo postal 48-3,
 62251 Cuernavaca, Mexico}
\affiliation{$^{3}$Center for Studies in Statistical Mechanics and Complex Systems, The
University of Texas at Austin, Austin, Texas 78712, USA}
\date{\today}

\begin{abstract}
We study the classical electron scattering from a driven inverted Gaussian
potential, an open system, in terms of its chaotic invariant set. This chaotic invariant set
is described by a ternary horseshoe construction on an appropriate Poincare
surface of section. We find the development parameters that describe the
hyperbolic component of the chaotic invariant set. In addition, we show
that the hierarchical structure of the fractal 
set of singularities of the scattering functions is the same as the structure
of the chaotic invariant set. Finally, we construct a symbolic encoding of the
hierarchical structure of the set of singularities of the scattering functions
and use concepts from the thermodynamical formalism to obtain one of the measures of chaos of the
fractal set of singularities, the topological entropy.
\end{abstract}      
 
\maketitle

\vspace{2cm}
\section{Introduction}
Simple one-dimensional atomic potentials in external time-periodic 
electric fields
have been used to predict several phenomena in the theory of laser-atom
interactions at high laser intensity such as stabilization with 
increasing laser intensity. These models are of 
particular interest because
their classical versions display chaotic motion \cite{Reichl}, thus 
providing insight into quantum-classical correspondence.
 
 The one-dimensional inverted Gaussian potential in the presence of
a strong time-periodic electric field has already offered interesting 
insights into different aspects of the laser-atom interactions 
\cite{Bardsley,Yao,Marinescu}. This short-range driven atomic
potential has also been used to study the phase-space picture of 
resonance creation and to show that the resonance states are scarred on 
unstable periodic orbits of the classical motion \cite{Timberlake}.  
 In addition, two of the authors have studied electron scattering from 
the driven inverted Gaussian and, using Floquet theory, they
constructed the Floquet scattering matrix. They found that the eigenphases
of the Floquet scattering matrix undergo a number of "avoiding crossings" as
a function of the electron Floquet energy \cite{Agapi} which is a quantum
manifestation of the destruction of the constants
of motion and the onset of chaos in classical phase space.    
  These "avoided crossings" were the motivating factor for a 
detailed study of the classical chaotic electron
scattering from the driven inverted Gaussian potential which is the 
focus of the current work. 
 
Of primary importance in chaotic scattering \cite{generalchaotic} is 
the identification of universal features which distinguish it from 
regular scattering.  For open systems, one such feature is the fractal set of singularities
observed in scattering functions such as the time-delay function
\cite{singularities}. This fractal set of singularities is the
result of the intersection of the incoming electron asymptotes 
with the invariant manifolds of the chaotic invariant set in the
asymptotic region. The chaotic
invariant set underlies the structure of the classical phase space in the
sense that its properties determine the quantities which characterise the
scattering process. One such property is the hierarchical structure of the 
chaotic invariant set which is the same as the structure of the fractal set 
of singularities of the time delay function.

In
section IIIA of this paper,    
 we obtain the hierarchical structure of the chaotic invariant set for 
the driven inverted Gaussian, which is an open system. The chaotic invariant
set is represented as a horseshoe
construction in an appropriate Poincare surface of section. We also obtain the
development parameter of the horseshoe construction which describes 
the hyperbolic component of the invariant set while it ignores non-hyperbolic
effects. In section IIIB we compute the time delay function and show that it
has a fractal set of singularities with the same structure as the hierarchical
structure of the invariant set. 
 In section IIIC we obtain a symbolic dynamics
\cite{symbolic1,symbolic6}, that is, a symbolic encoding of the branching tree,
 that describes the hierarchical structure of the chaotic invariant set and
thus the hierarchical structure of the fractal set of singularities of the
time delay function. Finally, using concepts from the thermodynamical
formalism \cite{Beck,Tel,Feigenbaum}, we obtain one of the measures of chaos
of the fractal set of singularities of the scattering functions, the
topological entropy.

\section{Model}
We study the classical scattering of an electron from a one-dimensional
inverted Gaussian atomic potential in the presence of a strong time-periodic
electric field. The electric field $E(t)=E_{0}sin(\omega t)$ ($T=2\pi/\omega$
is the period of the field) is treated within the dipole approximation as a 
monochromatic infinite plane wave linearly polarized along the direction of
the incident electron. In what follows, we work in the Kramers-Henneberger
(KH) \cite{Kramers,Henneberger} frame of reference, which oscillates with a free electron in the
time-periodic field. In the KH frame there are well defined asymptotic
regions where the electron is under free motion.
 The Hamiltonian in one space dimension $x$ that
describes the dynamics of the system in the KH frame is in atomic units (a.u.) \cite{Agapi}
\begin{equation}
\label{eq:classicalH1}
H(x,t)=\frac{p^2}{2}-V_{0}e^{-((x+\alpha(t))/\delta)^2},
\end{equation}
where $\alpha(t)=\alpha_{0}sin(\omega t)$ is the classical displacement
of a free electron from its center of oscillation in the time-periodic 
electric field $E(t)$ with $\alpha_{0}=-q E_{0}/\omega^2$ ($q$ is the
particle charge which for the electron is $q=-1$ a.u.).
Next, we transform Eq.(\ref{eq:classicalH1}) to a two-dimensional
time-independent system, where the total energy $E$ of the system is
conserved, as follows: 
\begin{equation}
\label{eq:classicalH}
H=\frac{p^2}{2}-V_{0}e^{-((x+\alpha_{0}\sin(\phi))/\delta)^2}+\omega I.
\end{equation}
$I$ and $\phi$ are respectively the action-angle variables 
of the driving field and $\phi=\omega t$. 
 In the limit $x\rightarrow \pm \infty$ the Gaussian atomic potential 
tends to zero faster than $1/x$ \cite{Agapi}. Thus, there are well defined asymptotic regions where
the electron is under free motion and its dynamics is described by the asymptotic 
Hamiltonian:
\begin{equation}
\label{eq:classicalHas}
H_{as}=\frac{p^2}{2}+\omega I.
\end{equation}
In the asymptotic regime, Eq.(\ref{eq:classicalHas}), the electron momentum, $p$, as 
well as the action of the field, $I$, are conserved quantities.
 In the following sections, all our calculations are performed
with the values $V_{0}=0.27035$ a.u. and $\delta=2$ a.u. assigned
to the parameters of the inverted Gaussian potential. These values 
of the parameters $V_{0}$ and $\delta$ were shown to describe well
the quantum behaviour of a one-dimensional model negative chlorine ion
$Cl^{-}$ in the presence of a laser field \cite{Yao,Marinescu,Agapi,Fearnside}. The frequency
of the time periodic field, $\omega$, and the amplitude of the field, $\alpha_{0}$, 
are taken constant and equal to $0.65$ a.u. and $0.9$ a.u., respectively.
 These values for the frequency and amplitude of the field were chosen
so that the resulting horseshoe construction is not prohibitively complicated to study.

\section{Chaotic scattering}

We are interested in understanding the underlying structure of the classical
chaotic scattering system under consideration. That implies knowledge
of the chaotic invariant set.
 In what follows, we first show how to
construct the hierarchical structure of the chaotic invariant set 
for the inverted Gaussian atomic potential
driven by a laser field. The scheme we follow to construct 
the hierarchical
structure of the chaotic invariant set is valid only for systems with two 
degrees of freedom. Then, we show how the structure of the chaotic 
invariant set allows us to understand the structure of the fractal set of singularities 
of the scattering functions. We then obtain a symbolic dynamics for the
hierarchical structure of the chaotic invariant set. This symbolic dynamics
describes the hierarchical structure of the scattering functions as well.
 We express this symbolic dynamics in the form of a
transfer matrix \cite{Feigenbaum} and compute the topological entropy
which in our case is a measure of the fractal structure of singularities of
the scattering functions.

\subsection{Chaotic invariant set}
The chaotic invariant set is usually represented by a horseshoe construction
in an appropriate Poincare surface of section. In the case of the well-known Smale horseshoe 
the construction is done by stretching a fundamental region $R$ and 
folding it on to the original region \cite{Smale,Ott}. The boundaries of $R$ are given by
segments of the invariant manifolds of the outer fixed points of the system.
 Following the above general scheme we first define the fundamental region
$R$.
 The system under consideration has three period-one periodic orbits (fixed
points). The inner fixed point is an elliptic one. The two outer fixed points
are located at $x\rightarrow \pm \infty$. As $x\rightarrow \infty$ the invariant 
stable and unstable manifolds of the outer fixed point $C$, see Fig.(\ref{fig:farea}), converge to the same
manifold (eigenvector), with $p=0$. The same is true for the manifolds of the fixed point
$A$ at $x\rightarrow -\infty$. So, globally, the outer fixed points behave as unstable ones,
that is, they produce invariant manifolds of the same topology as the one produced
by hyperbolic fixed points. However, in a small
neighbourhood around them they behave as parabolic ones. That is, the tangent map at $x\rightarrow \pm \infty$ has a degenerate eigenvalue equal 
to one (one eigenvector) \cite{Reichl}. The invariant manifolds of these outer fixed 
points determine the boundaries of the fundamental area $R$, see
Fig.(\ref{fig:farea}).

 In Fig.(\ref{fig:farea}) the horseshoe is constructed on the Poincare surface
of section
$\phi=\pi/2$. We use the Poincare surface of section $\phi=\pi/2$ for all our
calculations. This choice of the Poincare surface of section simplifies the horseshoe
construction because on this plane the time reversal transformation $t\rightarrow -t$ is
equivalent to the $p\rightarrow -p$ transformation. Thus, from the stable manifolds
of the outer fixed points one obtains the unstable manifolds by letting
$p\rightarrow -p$ and vice versa.
 The driven inverted
Gaussian has no right/left symmetry. That is, the Hamiltonian is not
invariant under the transformation $x\rightarrow -x$. Thus, the invariant
set of the system is described by a ternary (three fixed points) asymmetric horseshoe
construction. That is, the underlying structure of the scattering functions
for electrons incident from the right/left is described by two different views
right/left of the same horseshoe construction. The reason we consider the
invariant manifolds of the outer fixed points is that these are the manifolds
that are "seen" by the scattering trajectories and thus have an effect on the
scattering functions.
   
Let us now obtain the right view of the hierarchical structure of the horseshoe construction 
that underlies scattering for electrons incident from the right.
 The fundamental area $R$, see Fig.(\ref{fig:farea}), is defined by the
zero order tendrils as well as an infinite number of preimages/images
of the unstable/stable invariant manifolds, respectively.  
 We now add one iteration step
of the stable manifolds. That is, using Hamilton's equations of motion 
for the Hamiltonian given in Eq.(\ref{eq:classicalH}) we propagate
the points on the segments of the stable manifolds, $AD$ and $EC$ in 
Fig.(\ref{fig:farea}), backwards in time
for one period of the driving field (To obtain the tendrils of the unstable
 manifolds we propagate forward in time). The intersection of the first
image, first order tendrils, of the stable manifolds with the unstable manifold of the
fixed point $C$, segment $CD$ in
 Fig.(\ref{fig:farea}), reveals the first order gap $G_{1}^{s}$, see
 Fig.(\ref{fig:rightshoe}). The intersection with the unstable manifold of one more iteration
step of the stable manifolds reveals the second order gaps $G_{2}^{s}$, see
 Fig.(\ref{fig:rightshoe}). Thus, the gap $G_{n}^{s}$ is the area enclosed by
the nth order tendril of the stable manifold and the boundary of the fundamental area $R$.
 A point that lies in $G_{n}^{s}$ is mapped out of the fundamental region after $n$ applications of
the map, it is thus of hierarchy level $n$. These gaps play an important role
because they are areas which are not needed to cover the invariant set. No
higher level tendrils of the invariant manifolds will ever enter such gaps. 
 So, with each iteration step one further tendril of the stable manifolds is added and
one further level of hierarchy of these gaps is displayed \cite{Jung1}. We therefore
see the construction scheme of the horseshoe by going from one level of 
hierarchy to the next. We note that the term gaps corresponds to what is known as
lobes in fluid transport problems \cite{Wiggins}. In particular, the gaps
correspond to those lobes that are inside the area $R$.
 In a similar way, we construct the left view of the hierarchical structure of the horseshoe construction 
that underlies scattering for electrons incident from the left, 
 see Fig.(\ref{fig:rightshoe}). 
The 
intersection points of the stable manifolds with the unstable
 manifolds of the outer fixed points, seen in Fig.(\ref{fig:rightshoe})
 are the so called homoclinic/heteroclinic points 
for intersecting manifolds corresponding to the same
 (homoclinic) or different (heteroclinic) fixed points.
  These homoclinic/heteroclinic 
intersections underly the classical chaotic scattering.

Next, we compute the so called development parameter that approximately gives the 
development stage of the horseshoe construction. The significance of this 
parameter is that it describes universal aspects
of the horseshoe and ignores the details. That is, it determines the hyperbolic
component of the invariant set which is the important part for the scattering
behaviour and neglects non-hyperbolic effects that are due to the 
Kolmogorov-Arnold-Moser (KAM)
tori \cite{Jung1,Jung3,Jung4}. The non-hyperbolic effects appear at high levels of the hierarchy
as tangencies, non transversal intersections, between stable and unstable manifolds and have a very small effect on 
the scattering functions (see \cite{symbolic6} for more details on tangencies
between stable and unstable manifolds). For the values of the frequency and the amplitude
of the driving field we choose, there are tangencies when $4$th, $n=4$, order
tendrils of the stable manifolds are intersecting 4th order tendrils of
the unstable manifolds in the interior of the fundamental region. The effect
of these tangencies in the interior of the fundamental region becomes visible
in the scattering functions at a hierarchical level $2n$, in our case 8. The
reason is that if an $n$th order tendril of the stable manifold
intersects tangentially an $n$th order tendril of the unstable manifold in the
interior of the fundamental region, then the $n+1$ tendril of the stable
manifold will intersect the $n-1$ tendril of the unstable manifold, and so
on, until the $n+n$ tendril of the stable manifold intersects the zero
order tendril of the unstable manifold, that is, when
the $2n$ tendril of the stable manifold intersects the local segment (zero
order tendril) of the unstable manifold. But, as we show in the next section,
 it is exactly the structure of the intersections of the stable manifolds with
the local segment of the unstable manifold that is "picked" by the
scattering functions.

 The development parameter has the value $1$ for a
complete horseshoe. A horseshoe is complete when the 
tendril of level $1$ of the unstable manifold reaches the other side of the
fundamental area $R$. 
 For an incomplete horseshoe the development
parameter is determined by the relative length of the tendril of level $1$ of
the unstable manifold as compared to the complete case.
 It is given by $r_{n}N^{-n}$ \cite{Jung1}, where $n$ is the highest level of hierarchy
considered, $r_{n}$ is the number of the gap that the 
tendril of order $1$ of 
the unstable manifold reaches up to, counting the gaps starting from the fixed
point and N is the number of the fixed points.
 For the system under consideration $N=3$. It is important to realize that
the numbers are assigned to the gaps of the incomplete horseshoe construction
after comparing with the gaps of the complete horseshoe construction
\cite{Jung1}. Note, that the value of the formal parameter, given by
$r_{n}N^{-n}$, remains the same when different hierarchy levels are
considered. The reason is, that as we go from a hierarchy level $n$ to
the next hierarchy level $n+1$, $N-1$ gaps are added between successive 
gaps at the hierarchy level $n$, in the complete horseshoe construction.
 Thus, one can show that if the number $r_{n}$ is assigned to a certain gap at hierarchy level
$n$, the number $r_{n+1}$ assigned to the same gap at hierarchy level $n+1$
is $r_{n+1}=Nr_{n}$. So, $r_{n+1}N^{-(n+1)}=r_{n}N^{-n}$ and the value of the
formal parameter remains the same.   
 
As already mentioned, for $\omega=0.65$ a.u. and $\alpha_{0}=0.9$ a.u.
the driven inverted Gaussian is described by a ternary asymmetric horseshoe
construction and it is thus described by two development parameters.
 The development parameter that corresponds to the manifolds of the fixed
point at $x\rightarrow -\infty$, $A$, 
has the value $1$ since the first order tendril of the unstable manifold of
the fixed point $A$ reaches the other side of the fundamental area $R$, see 
Fig.(\ref{fig:rightshoe}).
 The development parameter that corresponds to the manifolds of the fixed
point at $x\rightarrow +\infty$, $C$, has the value $1/3$ as can be seen in 
Fig.(\ref{fig:rightshoe}). The value $1/3$ is obtained as follows: if we
consider tendrils up to hierarchy level $n=1$ then the first order tendril 
of the unstable manifold of the fixed point $C$, $t_{1}^{u,C}$, reaches up to the $r=1$ gap.
 If we consider tendrils up to hierarchy level $n=2$ then $t_{1}^{u,C}$ reaches
up to the $r=3$ gap and for hierarchy level $n=3$ $t_{1}^{u,C}$ reaches up to the $r=9$ gap. That is, the value of the development parameter remains
the same when different hierarchy levels are considered.
  In Fig.(\ref{fig:KAM}), we see how the KAM
tori around the middle fixed point cause an incomplete horseshoe
construction. 
 So for $\omega=0.65$ a.u. and $\alpha_{0}=0.9$ a.u. the chaotic invariant set
is described by a ternary asymmetric horseshoe construction with development
parameters $1$ and $1/3$. For reasons explained at the end of section II, the
frequency is taken equal to 0.65 a.u. (high frequency regime compared to
$V_{0}=0.27035$ a.u.). For this
frequency a horseshoe with development parameters $1$ and $1/3$ is realized 
approximately in the interval $(0.7,1.15)$ a.u. of the amplitude of the field,
 $\alpha_{0}$.

\subsection{Scattering functions}

 The scattering functions 
give properties of the final electron asymptotes as a function
of the incoming electron asymptotes. In the case of classical chaotic scattering
the scattering functions have a fractal set of 
singularities. This fractal set of singularities is the result of 
the intersection of the incoming electron asymptotes with the 
underlying chaotic invariant set. That is, when the scattering electron
trajectory starts exactly on the stable manifold of the chaotic invariant set
it stays on the chaotic set forever, resulting in a singularity of the
scattering
function.
 Furthermore, the structure of the 
set of singularities is the same as the structure of the chaotic invariant
set \cite{Jung1}.

In what follows, we compute the time delay, $T^{del}$, one of the most important scattering 
functions. The time delay is a measure of how much the  
incoming electron delays due to its interaction with the potential
in the scattering region and is given by:
 \begin{equation}
\label{eq:timedelay}
T^{del}=T-
|\frac{x_{in}}{p_{in}}|-|\frac{x_{out}}{p_{out}}|.
\end{equation}
$T$ is the time it takes for the electron to travel from
the incoming to the outgoing asymptotic region. There is an 
arbitrariness in the time $T$ due to the specific choice of the
initial distance $x_{in}$ that the timing is initiated in
the incoming asymptotic region and the final
distance $x_{out}$ that the timing is stopped in the outgoing
asymptotic region. To remove this 
arbitrariness we substract the time that the electron spends running
along the initial and final asymptotes, $|\frac{x_{in}}{p_{in}}|$ and
$|\frac{x_{out}}{p_{out}}|$ respectively.

We consider scattering from the right and compute the time delay function
for a line of initial conditions
 in the asymptotic regime that completely intersects one tendril of 
the stable manifold of the outer fixed point $A$, see Fig.(\ref{fig:images}). 
We compute the time delay function, for the choice of initial conditions
denoted as $0$ in Fig.(\ref{fig:images}), as a
function of the initial momentum, $p_{in}$, along the line of initial
conditions, see Fig.(\ref{fig:Timedelay}). This choice of initial
conditions allows us to understand the structure of singularities of the
time delay function as follows. From Fig.(\ref{fig:images}) we see that the
iterates in time of the
line of initial conditions converge toward the boundary of the fundamental
region that is defined by the local segment of the unstable   
manifold of the fixed point $C$.
 The intersections of the line of initial conditions with the
stable manifold of the fixed point $A$ are mapped on 
intersections of the iterates with
the same stable manifold. Thus, the singularity structure of the scattering
function is the same as the pattern resulting from the intersection of the 
stable manifolds with the local segment of the unstable manifold of the fixed
point $C$.
 That implies that the intervals of continuity of the scattering function correspond     
to the gaps that the tendrils of the stable manifolds cut into the
 fundamental area of the horseshoe construction. In other words, the pattern
of the fractal set of singularities of the time delay function is the same
as the hierarchical structure of the horseshoe construction. 
 We further illustrate this point as follows.  
 In 
Fig.(6a), we compute  
 the hierarchy level of the intervals of continuity for part of the time delay
function, see Fig.(6b) (Fig.(6b) is a magnification of part of Fig.(\ref{fig:Timedelay})). 
   To do so, we initiate trajectories at the intervals of continuity of the delay
function and count 
the number of times the scattering trajectories "step" into the fundamental
region, see Fig.(6a). If a scattering trajectory "steps" inside the area $R$ $n-1$ times
that means that it takes $n$ applications of the map before it is mapped
outside of $R$. We thus say that the trajectory was initiated at an interval of continuity
of hierarchy level $n$. For example, we see from Fig.(6a) that the scattering
trajectory with $p_{in}=-0.1973$ "steps" two times inside $R$. Thus, the
interval of continuity it was initiated at is of hierarchy level three.
  The resulting pattern of singularities shown in Fig.(6a) is the same 
as the pattern of singularities of the time delay function as a comparison of 
Figs.(6a) and (6b) reveals. 

 Let us now explain how the hierarchy level of the
intervals of continuity is related to the gaps of the horseshoe
construction. As we illustrate in Fig.(\ref{fig:map}), if a scattering
trajectory approaches the local segment of the unstable manifold along a gap
of order $n$, then it steps inside the area $R$ $n-1$ times before it is
mapped outside $R$. At the same time, if the scattering trajectory steps
inside the area $R$ $n-1$ times that means that it is mapped outside of $R$ after
$n$ applications of the map and thus the trajectory was initiated at
an interval of continuity of hierarchy level $n$. Thus, a gap of hierarchy
level $n$ of the horseshoe construction corresponds to an interval of
continuity of hierarchy level $n$ of the time delay function. That implies
that the hierarchical structure of the chaotic invariant
set and of the scattering functions is the same. Indeed, a comparison of Figs.(6a) and (\ref{fig:branchright})
(Fig.(\ref{fig:branchright}) is explained in the following section) reveals that the
pattern of singularities of the time delay function in Fig.(6b) is
the same as that part of the hierarchical structure of the chaotic invariant
set that is encircled by a square in Fig.(\ref{fig:branchright}).

 For the system under 
consideration the potential in the interaction region is known and so
we can directly obtain the hierarchical structure of the chaotic invariant set
and thus the structure of the scattering functions. However, when the
potential in the interaction region is not known, then one has to find 
 from asymptotic observations the hierarchical structure of the
scattering functions in order to obtain the structure of the chaotic invariant
set.

\subsection{Measures of Chaos}

It is possible to construct a topological measure of the degree of chaos
contained in this
scattering system if we can construct a symbolic dynamics which reproduces
the hierarchy of
intersections of the stable and unstable manifolds.  The first step is to
obtain the
branching trees that describe the right/left view of the horseshoe constructions for scattering
from the right/left respectively. The second step involves the development of a symbolic
dynamics which
reproduces the structure of the branching trees. It is important to note
that for the
values of the amplitude and the frequency of the driving field considered
there are
tangencies between the stable and unstable manifolds on the 4th order tendrils.
These tangencies introduce non-hyperbolic effects that will cause a breakdown of the
symbolic dynamics
starting from hierarchy level $8$ and higher. However, knowledge of the
symbolic dynamics
up to hierarchy level 8 gives a significant measure of the degree of
observable chaos in
this scattering system.

\subsubsection{Branching Trees}

  Let us first  obtain a branching tree \cite{Jung1}, that describes the
right view of the horseshoe construction for scattering from the right.  We will use
information developed in
Section IIIA. First, let us  consider the interval $I_{1}^{0}$ which
corresponds to the local segment of the unstable manifold $CD$ of the
fixed point $C$
(see Fig.(\ref{fig:branch})). This is the first step in the construction of the branching
tree and corresponds to hierarchy level $n=0$. In the
second step, hierarchy level n=1, the first order tendril of the stable manifold
of the fixed point $A$  cuts the interval $(s_{0},s_{1})$ out of
$I_{1}^{0}$ and leaves
two intervals $I_{1}^{1}$ (the segment of $CD$ from $D$ to $s_{0}$) and
$I_{2}^{1}$ (the segment of $CD$ from $s_{1}$ to $C$). In the third step, hierarchy level $n=2$, the second
order tendril of the stable manifold of the fixed point $A$ cuts
the interval $(s_{4},s_{5})$ out of $I_{2}^{1}$ and leaves two intervals,
$I_{21}^{2}$ (the
segment of $CD$ from $s_{1}$ to $s_{4}$) and $I_{22}^{2}$ (the segment of
$CD$
from $s_{5}$ to $C$). In the same step (the same iteration) the second
order tendril of the stable manifold of the fixed point $C$ cuts
the interval $(s_{2},s_{3})$ out of $I_{1}^{1}$ and leaves two intervals,
$I_{11}^{2}$ (the segment of
$CD$ from $D$ to $s_{2}$) and $I_{12}^{2}$ (the segment of $CD$
from $s_{3}$ to $s_{0}$). Continuing this process we obtain the branching
tree shown in Fig.(\ref{fig:branchright}).

  In a similar way, we construct the branching tree that describes the
left view of the horseshoe
construction for scattering from the left, see Fig.(\ref{fig:branchleft}).
  The hierarchical structure of these branching trees is the same as the
hierarchical
structure of the chaotic invariant set.

\subsubsection{Symbolic Dynamics}

Having determined the geometry of the branching trees, we can now construct
a symbolic
dynamics that encodes the branching trees. In principle, since we have a
non-hyperbolic
horseshoe construction one needs an infinite number of grammatical
rules to construct a symbolic dynamics. However, we can construct an
approximate symbolic dynamics that describes well the outermost hyperbolic
component of the horseshoe construction. The symbolic encoding of the
branching tree is
not unique, but the measures of chaos one obtains for different encodings
are the same.

Our symbolic dynamics consists of four symbol values A,B,C and + and a set of
grammatical rules that allow us to encode each branch of the
branching tree. That is, each branch of the tree of hierarchy level n, is
labeled by a vertical sequence (string) of $n$ symbols made out of the four
symbol values A,B,C, and +. Each symbol sequence is read vertically up the branch of
the tree (see Figs.(\ref{fig:branchright}) and  (\ref{fig:branchleft})). The
order in which the four symbol values appear in each branch of the tree
is determined by the grammatical rules. That is, the rules tell us which of the four 
symbol values are allowed to be appended to a given branch of the tree as we go from
a certain hierarchy level to the next. 

Our rules depend on the last
"word" that appears on a given branch. This "word" is a vertical sequence of 
one two or three symbols and can be either of the eleven "words": A, ++, B+C,
 C+C, ++C, CC, BC, AC, B, B+ and C+ (see Figs.(\ref{fig:branchright}) and
(\ref{fig:branchleft})). The rules are:

\begin{itemize}
\item After a string (branch) ending in $A$,  $B+C$, $C+C$ or $++$ it is
allowed to attach the
symbols $A$, $B$ and $C$, going from left to right (standard
orientation). Thus, three strings (branches) stem out ending in $A$, $B$ and $C$.
\item After a string (branch) ending in $AC$, $B$, $BC$, $B+$, $C+$, $CC$ or
$++C$ it is
allowed to attach the
symbols $+$ and $C$, going from left to right (standard
orientation). Thus, two strings (branches) stem out ending in $+$, $C$.
\item $B$ always inverts the previous orientation.
\item $C$ always inverts the previous orientation if it comes after $S+$,
where $S$
is not $+$.
\item $+$ always inverts the previous orientation if it comes after $S$
where $S$ is
not $+$.
\end{itemize}

By {\it previous orientation} we mean the following: if at a hierarchy
level $n$
there are three branches ending, for example, in the symbols $A$, $B$ and
$C$, going from
left to right (see
Fig.(\ref{fig:branchleft})), then at hierarchy level $n+1$, from the
string ending in $B$ two strings stem out with
symbol endings $+$ and $C$, according to the second grammatical rule. According to the third grammatical rule,
the symbol endings $+$ and $C$, going from left to right,
 at level
$n+1$, must have the inverse orientation to the one of the symbol endings at
level $n$. In this example, at level $n$, the symbol endings $A$, $B$ and
$C$, going from left to right, have the standard orientation. Thus, after
the string ending in $B$, two branches stem out, at level $n+1$, with symbol endings $C$ and $+$, going from left
to right,  see Fig.(\ref{fig:branchleft}).

To symbolically encode the right/left branching trees
in Figs.(\ref{fig:branchright}) and (\ref{fig:branchleft}) we have started
at level $n=1$ by
attaching the symbols
$+$ and $C$ for the right and $A$, $B$ and
$C$ for the left view of the
branching trees, respectively, and then use the above grammatical rules to
continue the
encoding. Using these rules we can encode and thus obtain the structure of
the
branching trees safely up to hierarchy level 7. For the
values of the
frequency and amplitude of the driving field we consider here, there are
tangencies between
the invariant manifolds at level four in the interior of the fundamental
region. These tangencies can
cause our symbolic encoding to break down at hierarchy level 8 and higher of
the branching tree.
 That is, these tangencies can introduce additional branches in the branching 
tree, starting at level 8, which are not accounted for by our grammatical rules.  Note, that the above described symbolic dynamics encodes the branching
trees of the scattering functions as well.

If we now use concepts from a thermodynamical formalism \cite{Beck,Tel},
we can express the above described grammatical rules in the form of a transfer
matrix \cite{Feigenbaum}.
  To construct the transfer matrix
we use as entries the eleven "words" listed earlier. The
matrix element $(l,m)$ is $1$ if it is possible to attach to the "word" $l$
a symbol such that the resulting string ending is the "word" $m$, otherwise
the matrix element is $0$. In other words, if the transfer matrix
element
$(l,m)$ is one it means that if at a certain hierarchy level we have a
string
ending in the "word" $l$ when we go to the next hierarchy level it is allowed
to encounter a string ending in the "word" $m$.
 To clarify this
point,
consider for example the string ending with the "word" $l=++$ (see
Fig.(\ref{fig:matrix})).
According to the first grammatical rule, after the "word" $++$ we
can attach three symbols labeled $A$, $B$ and $C$ and so obtain the
strings $++A$, $++B$ and
$++C$. These strings have
the string endings, $m=A$, $m=B$ and $m=++C$, respectively, which can be
identified with three of the eleven "words".  Thus,
the
matrix elements $(++,A)$, $(++,B)$ and $(++,++C)$ are one, while all other
matrix
elements with $l=++$ are $0$.

Having constructed the transfer matrix, we can now compute the topological
entropy of
the branching tree. The topological entropy is a measure of the degree of 
chaos in the scattering system. Let us first describe the relation between the
topological entropy and
the transfer matrix.
  The topological entropy $K_{0}$ is the rate of exponential growth of the
number of intervals $Z(n)$, or equivalently the number of branches $Z(n)$,
at a
hierarchical level $n$ when $n$ is large with $Z(n)=e^{nK_{0}}$
\cite{Beck}. It directly follows that
$K_{0}=ln(Z(n+1)/Z(n))$. But, for large $n$, $Z(n+1)/Z(n)$ is
the average branching ratio of the trees. This ratio is given by the
largest eigenvalue of the transfer matrix \cite{Feigenbaum}.  For our
system, the largest
eigenvalue is
$\approx 2.31$. Thus, the topological entropy of the branching tree is
$K_{0}\approx 0.84$. This topological entropy describes the rate of growth
of the branches in the hierarchical structure of the scattering
functions and is thus a measure of chaos of the fractal set of
singularities.

It is useful to mention that for a horseshoe with $N$ fixed points
the value of the topological entropy, $K_{0}$, can vary between $0$ and
$ln(N)$. This is easily understood, since for a horseshoe with $N$ fixed
points the maximum value of the average branching ratio is $N$ and $K_{0}$
is the logarithm of the average branching ratio. Thus, for a ternary
horseshoe
construction, the case currently under consideration, $K_{0}$ can vary
between
$0$ and $ln(3)\approx1.1$. For the values of the frequency and amplitude
of
the driving field considered in this paper, we find that
$K_{0}\approx0.84$,
  close to the maximum value of $1.1$, which suggests that our system is in
the regime of strong chaos.

\section{Conclusions}

In this paper, we have studied the classical electron scattering from
a driven inverted Gaussian potential which is an open system. We have
shown
that the fractal pattern
of singularities of the scattering functions can be understood in terms of
the hierarchical structure of the chaotic invariant set which underlies
the
chaotic dynamics. We have constructed a symbolic encoding of the
hierarchical
structure of the chaotic invariant set. Using concepts from the
thermodynamical formalism, we have used this encoding to obtain the
topological entropy of the fractal set of singularities of the scattering
functions.

\newpage

\begin{figure}
\begin{centering}
\leavevmode
\epsfxsize=0.6\linewidth
\epsfbox{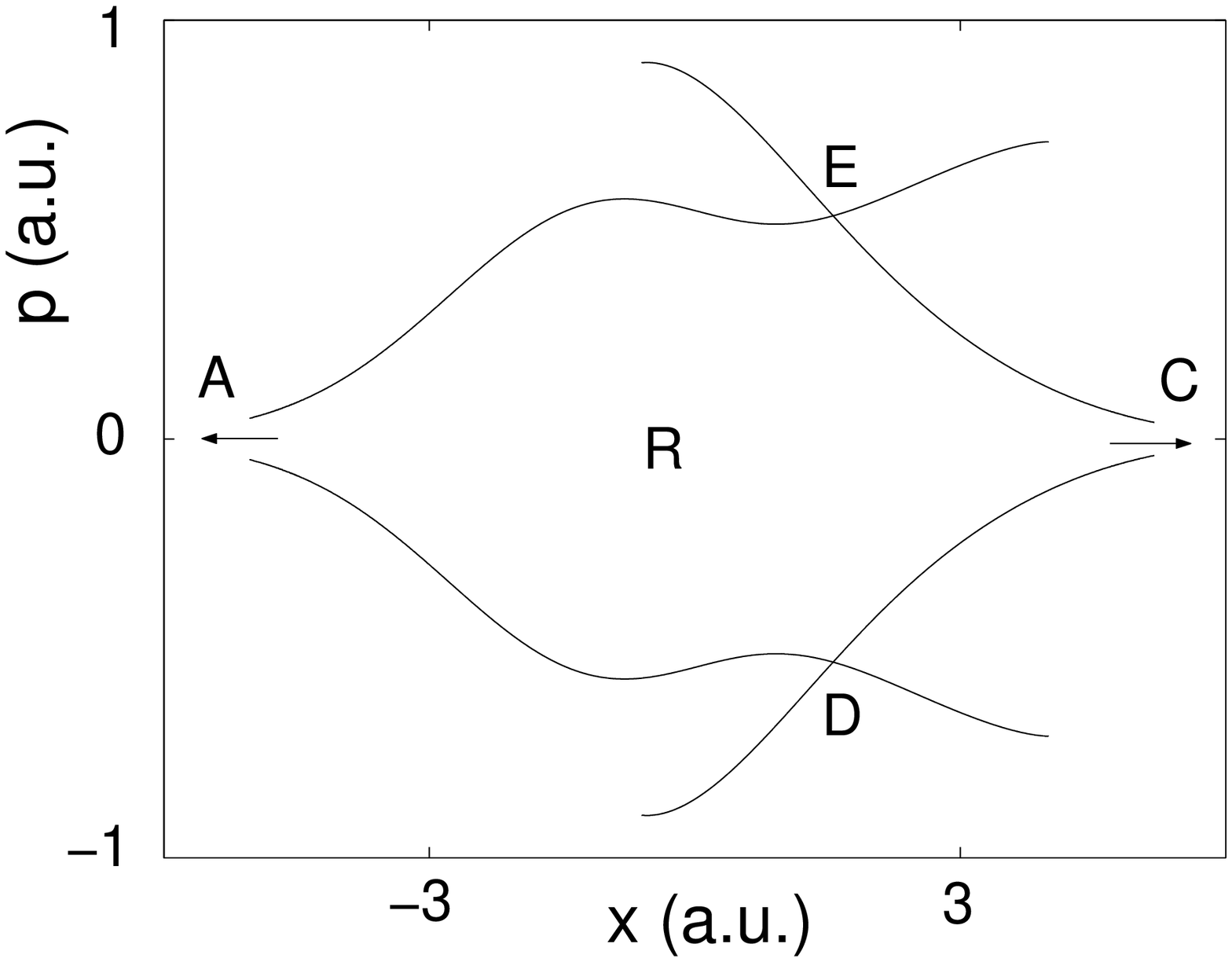}
\caption{}
\label{fig:farea}
\end{centering}
\end{figure}

\begin{figure}
\begin{centering}
\leavevmode
\epsfxsize=0.6\linewidth
\epsfbox{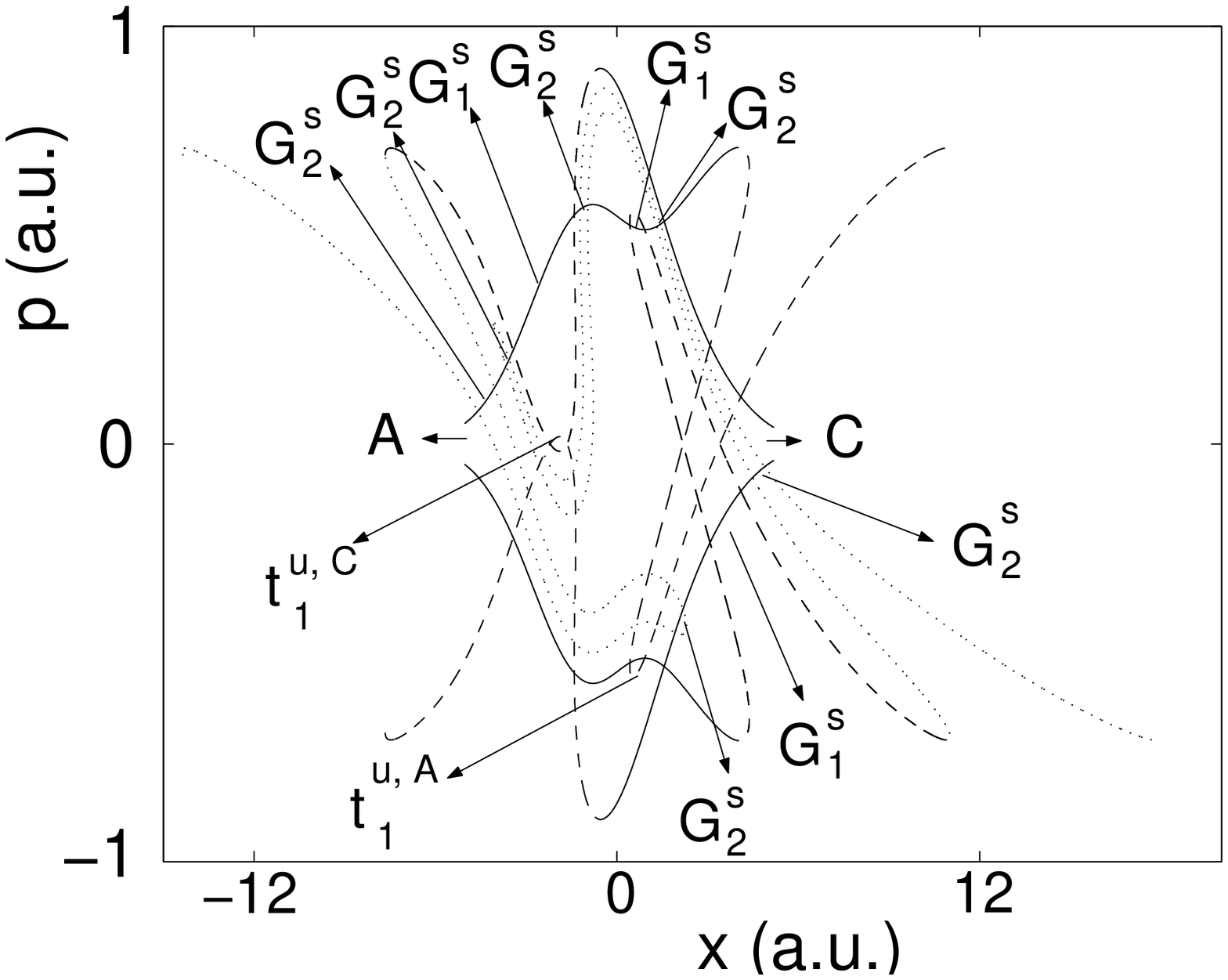}
\caption{}
\label{fig:rightshoe}
\end{centering}
\end{figure}

\begin{figure}
\begin{centering}
\leavevmode
\epsfxsize=0.6\linewidth
\epsfbox{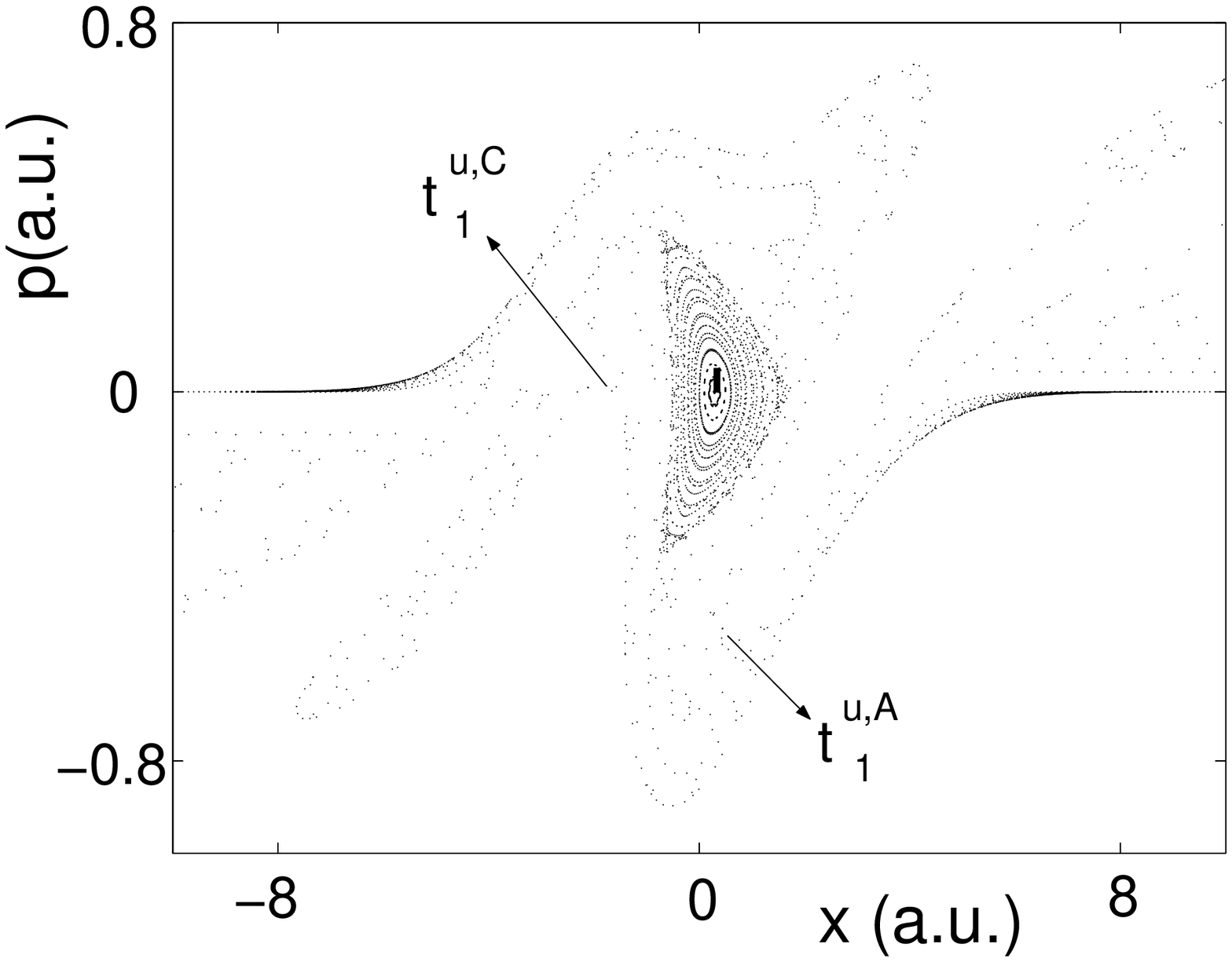}
\caption{}
\label{fig:KAM}
\end{centering}
\end{figure}

\begin{figure}
\begin{centering}
\leavevmode
\epsfxsize=0.6\linewidth
\epsfbox{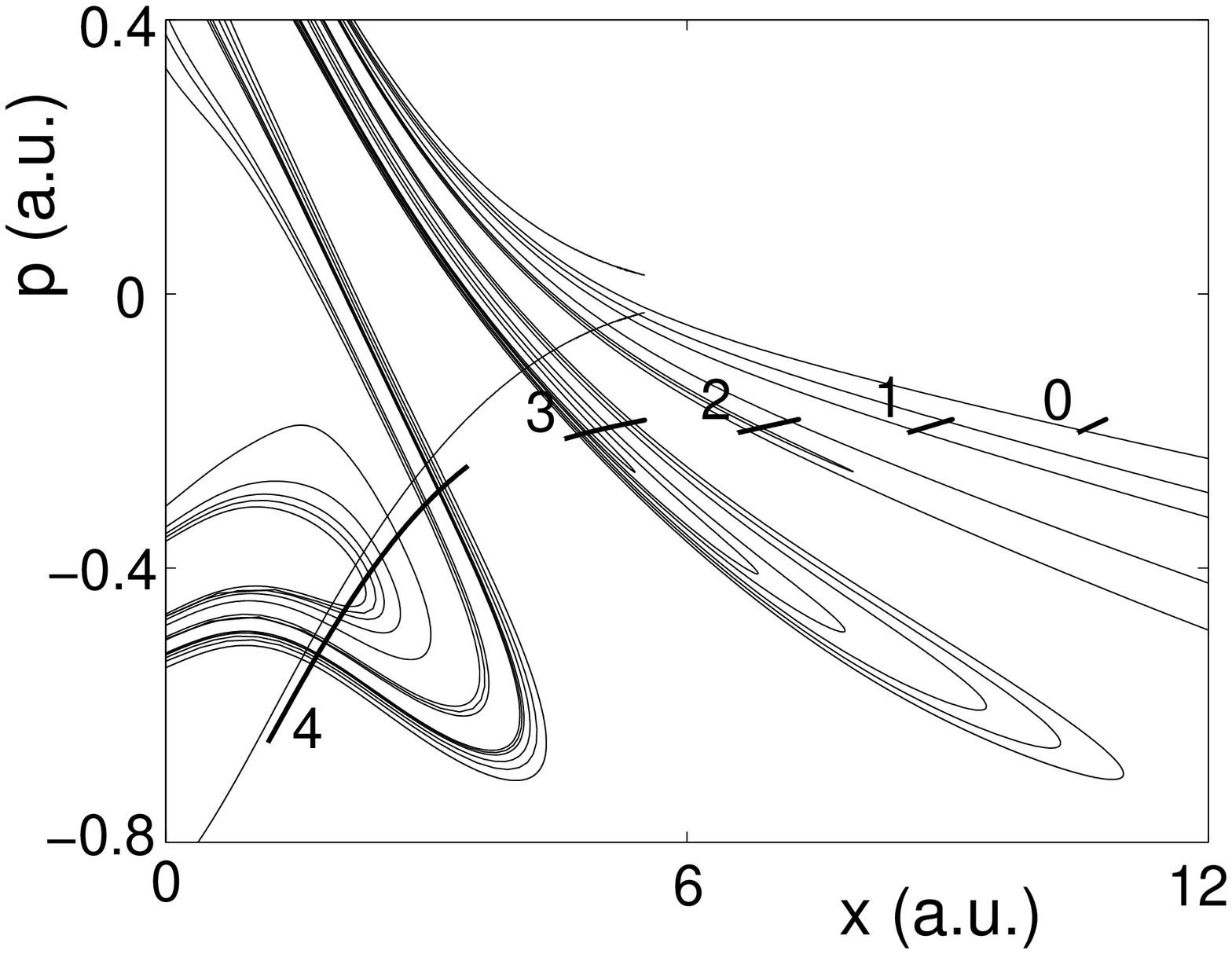}
\caption{}
\label{fig:images}
\end{centering}
\end{figure}

\begin{figure}
\begin{centering}
\leavevmode
\epsfxsize=0.6\linewidth
\epsfbox{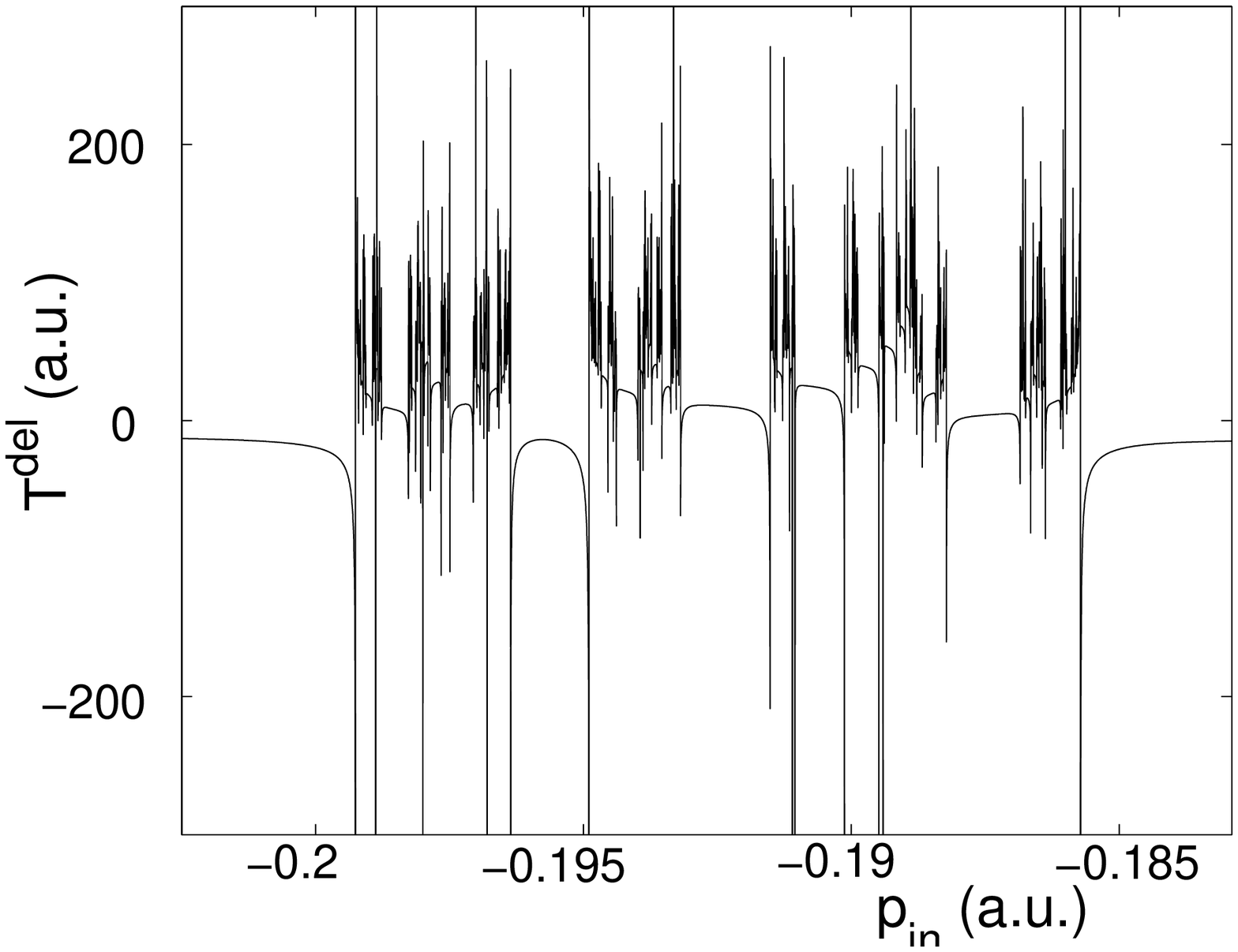}
\caption{}
\label{fig:Timedelay}
\end{centering}
\end{figure}

\begin{center}
List of Figures
\end{center}
\begin{itemize}
\item Fig1. The fundamental region $R$ is formed by the unstable manifold of the fixed point $A$, 
 segment $AE$, by the stable manifold of the fixed point $C$, segment $CE$,
 by the unstable manifold of the fixed point $C$, segment $CD$, and the stable
manifold of the fixed point $A$, segment $AD$.

\item Fig2. Horseshoe construction up to hierarchy level two on the Poincare surface of section $\phi=\pi/2$. The solid lines
indicate tendrils of order zero, the dashed lines indicate tendrils of order
one and the dotted lines indicate tendrils of order two. The gaps $G_{n}^{s}$ 
on the bottom right/top left  are formed by intersections of the stable manifolds of the
fixed points $A$ and $C$ with the local segment of the unstable manifold 
of the fixed point $C$/$A$, that is, $CD$/ $AE$. These intersections describe
the right view/left view
of the horseshoe construction. $t_{1}^{u,A}$
indicates the first order tendril of the unstable manifold of the fixed point
$A$. $t_{1}^{u,C}$
indicates the first order tendril of the unstable manifold of the fixed point
$C$.

\item Fig3. The initial conditions used to generate this strobe plot lie
on the line $p=0$. This strobe plot is generated by evolving the trajectories
forward in time and it thus "picks" the unstable manifolds of the fixed points
$A$ and $C$. The location of the middle fixed point, $B$, (period-1 orbit)
is located at $x=0.29$ and is indicated by a filled rectangle. Comparing
with Fig.(\ref{fig:farea}), we see that the first order tendril of the
unstable manifold of the fixed point $A$, $t_{1}^{u,A}$, penetrates  
the fundamental area $R$ completely. In the case though of the first order tendril of the
unstable manifold of the fixed point $C$, $t_{1}^{u,C}$, the KAM tori 
around the fixed 
point $B$ prevents it from reaching the boundary of the fundamental area $R$.

\item Fig4. For scattering from the right, we indicate as $0$ the line of
initial conditions in the asymptotic region used to compute the time 
delay function. This set of initial
conditions intersects the stable manifold of the fixed point $A$.
 The numbers $1-4$ indicate successive iterations in time of the set of initial conditions.

\item Fig5. Time delay function as a function of the initial momentum for the
set of initial conditions shown in Fig.(\ref{fig:images}).

\item Fig6. In a) we show the hierarchy level of the intervals of
continuity for part of the time delay 
function, see 
Fig.(\ref{fig:Timedelay}). For a given pair of initial values, $x_{0},p_{0}$, we propagate the
trajectories until they reach one of the asymptotic regions and count the
number of times the trajectory steps in the fundamental area $R$. In b)
we plot the time delay function for the same range of initial conditions as for
the hierarchy level of the intervals of continuity shown in a). We can
immediately see that 
both functions have the same pattern of singularities.

\item Fig7. The solid lines indicate tendrils of order zero, the dashed lines
tendrils of order one, the dotted lines tendrils of order two and the
dashed-dot line tendrils of order three.
 We initiate a trajectory in the right asymptotic region with
$p_{in}=-0.1973$ which is inside an
interval of continuity, see Fig.(6a). We then successively
iterate the trajectory in time (stars). The successive iterations are indicated by
numbers 1-8, respectively. The trajectory approaches the
local segment $CD$ of the unstable manifold of the fixed point $C$ inside the third order tendril 
of the stable manifold of the fixed point $A$ along a third order gap. One more iteration in time
maps area $a$ (shaded by dots), which is enclosed by the third order
tendril of the stable manifold of the fixed point $A$ and its unstable manifold,
 into area $b$ (shaded
by lines), which is enclosed by the second order tendril of the stable
manifold of point $A$ and its unstable manifold. 
 A further iteration in time maps area $b$ into area $c$ (shaded by lines),
 which is enclosed by the first order tendril of the stable manifold of point $A$ and its
unstable manifold. Finally, area $c$ is mapped to area $d$ (shaded by lines) and
enclosed by the zero order tendril of the stable manifold of point $A$ and its
unstable manifold. But area $d$ is outside the fundamental region and thus the
trajectory steps inside the fundamental region two times. Generally, if 
the scattering trajectory approaches the local segment of the unstable
manifold along a gap of hierarchical order n it will "step" inside the
fundamental area, $R$, n-1 times before it is mapped outside $R$.

\item Fig8. Construction of the branching tree for scattering from the
right. The first order gap $G_{1}^{s}$ reduces the initial interval $I_{1}^{0}$,
 at hierarchy level $n=0$, down to the two intervals $I_{1}^{1}$ and
$I_{2}^{1}$. Note that for the scattering functions, we obtain 
exactly the same branching tree as for the chaotic invariant set. 
 For the scattering functions, instead of the gaps it is the intervals of
continuity that are cut out from the original interval in a Cantor set
structure. 

\item Fig9. Branching tree and symbolic dynamics for scattering from the
right shown up to hierarchy level four. Each interval corresponds to one
branch of the tree. A branch at
hierarchy level $n$ is described by a string of length $n$.

\item Fig10. Branching tree and symbolic dynamics for scattering from the
left shown up to hierarchy level four. Let us now explain what we mean by
previous orientation in terms for example of the third grammatical rule. For example, at the hierarchical level $n=2$, indicated
by the arrow, the symbol endings of the three branches going from left to right are
$A$, $B$ and $C$ which is what we define as standard orientation. After the symbol
value $B$ we can attach the symbols $+$ and $C$ at the hierarchical level $n=3$,
 indicated again by an arrow.
 According to the third grammatical rule the symbols $+$ and $C$, at level $n=3$,
will be attached, after $B$, so that they have the inverse orientation of the
symbol values at level $n=2$. So, since at level $n=2$ the symbols $A$, $B$ and $C$
are attached in standard orientation then at level $n=3$ we attach after $B$
symbols $C$ and $+$ going form left to right, resulting in an inverse orientation
compared to the one at level $n=2$. Thus, we say that $B$ always inverts the
previous orientation.

\item Fig11. Transfer matrix.
\end{itemize}

\newpage

\begin{figure}
\begin{centering}
\resizebox*{0.8\textwidth}{!}{\includegraphics{total.eps}}
\caption{}
\label{fig:hierarchy}
\end{centering}
\end{figure}

\begin{figure}
\begin{centering}
\leavevmode
\epsfxsize=1.\linewidth
\epsfbox{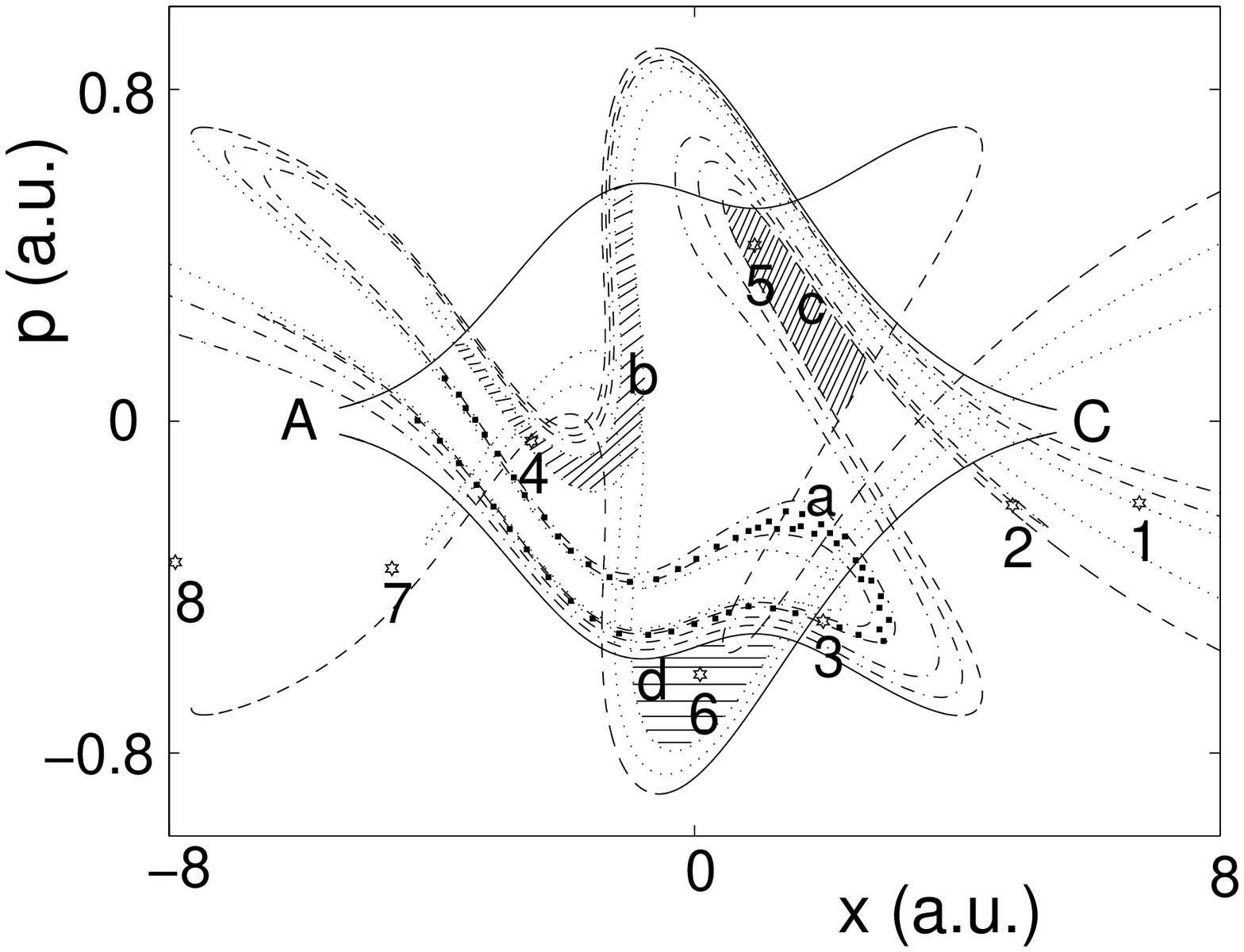}
\caption{}
\label{fig:map}
\end{centering}
\end{figure}

\begin{figure}
\begin{centering}
\leavevmode
\epsfxsize=0.6\linewidth
\epsfbox{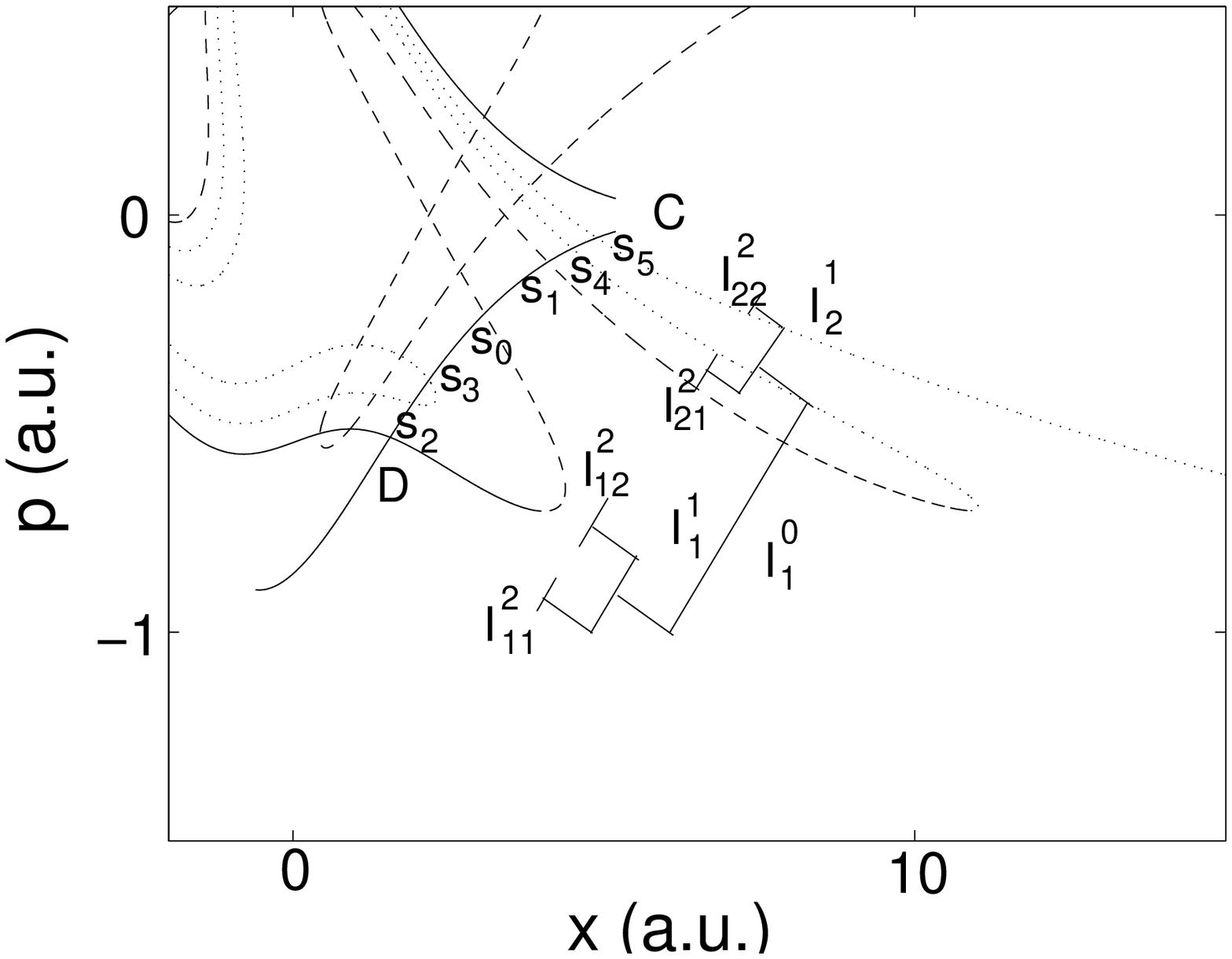}
\caption{}
\label{fig:branch}
\end{centering}
\end{figure}

\begin{figure}
\begin{centering}
\leavevmode
\epsfxsize=0.6\linewidth
\epsfbox{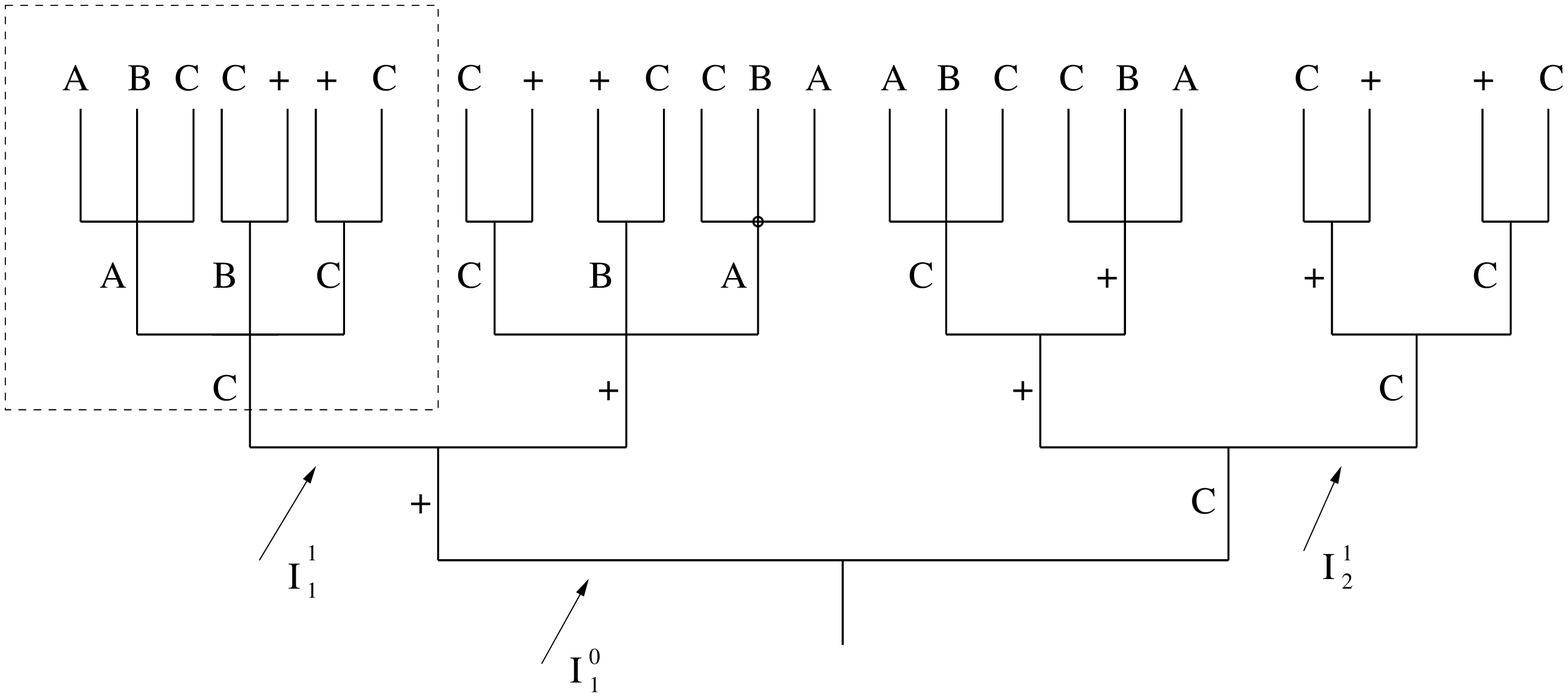}
\caption{}
\label{fig:branchright}
\end{centering}
\end{figure}

\begin{figure}
\begin{centering}
\leavevmode
\epsfxsize=0.6\linewidth
\epsfbox{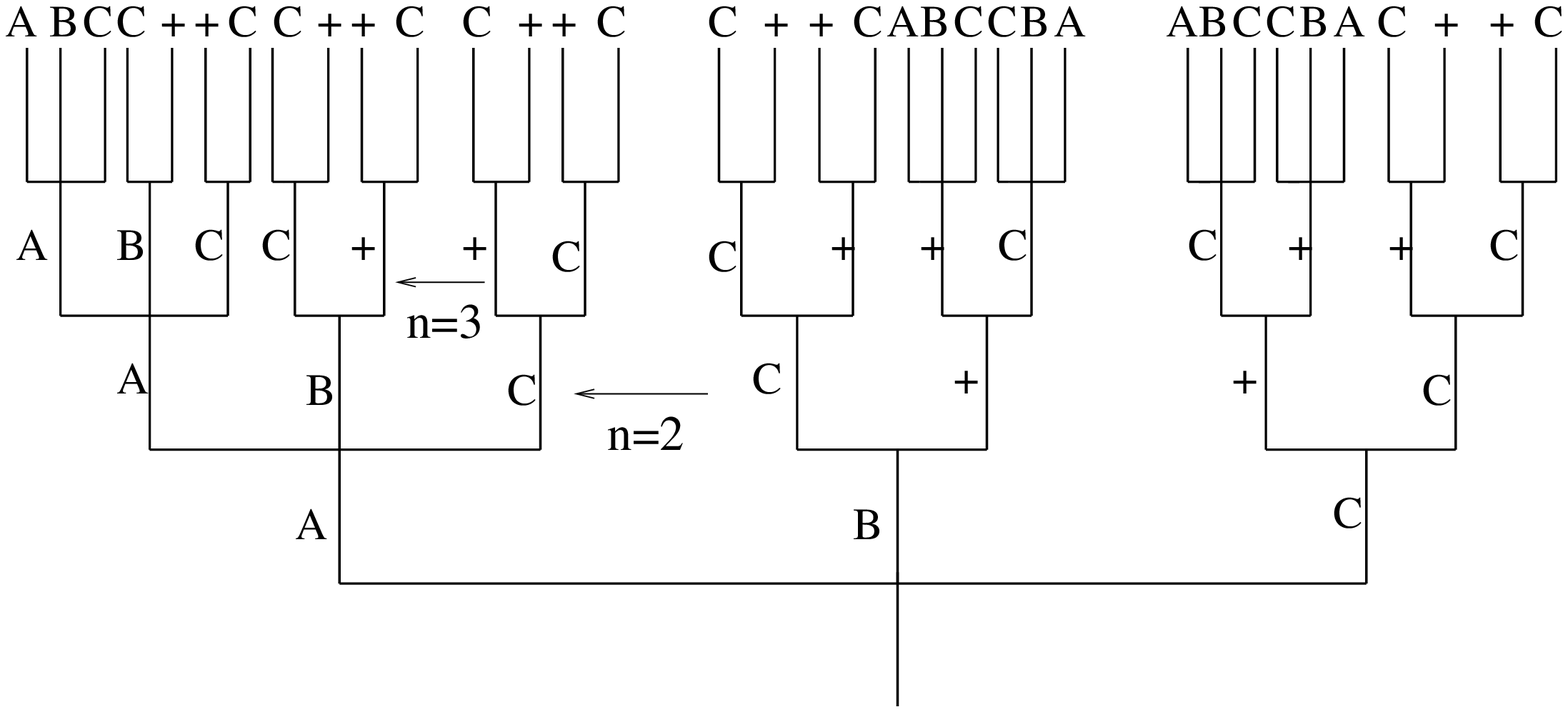}
\caption{}     
\label{fig:branchleft}
\end{centering}
\end{figure}

\begin{figure}
\begin{centering}
\leavevmode
\epsfxsize=0.6\linewidth
\epsfbox{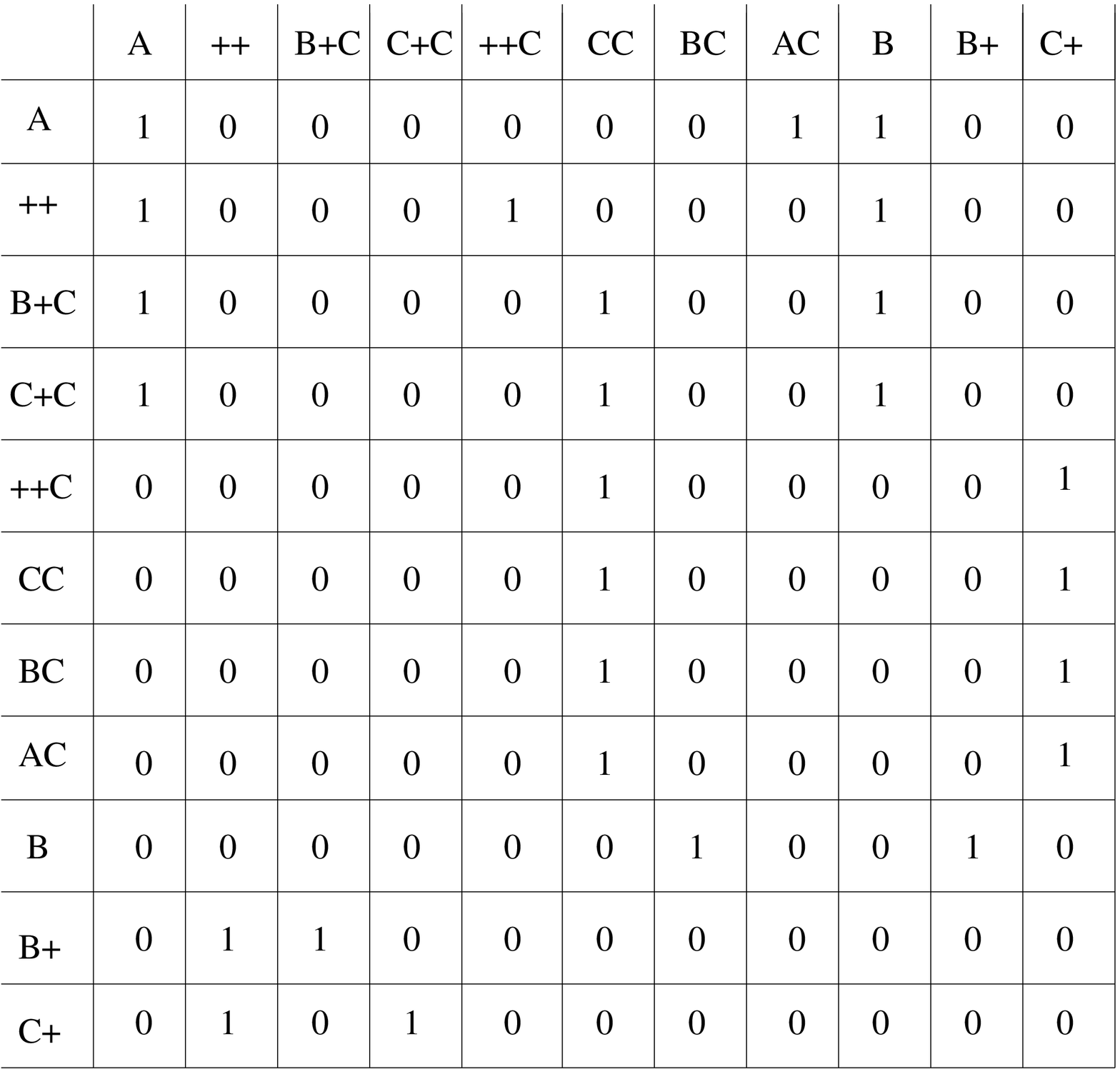}
\caption{}
\label{fig:matrix}
\end{centering}
\end{figure}

\end{document}